There are two style files :sty1, sty2 in front of the main text

\\ sty1.sty

\newif\ifnoncomplete

\def\final{\noncompletefalse\typeout{** FINAL form (substyle:Sachiko)}}
\noncompletefalse

\def \vol(#1,#2,#3){\ifrefphysrev{{\bf {#1}},
{#3} (19{#2})}\else{{{\bf {#1}} (19{#2}) {#3}}}\fi}
\def \NP(#1,#2,#3){Nucl.\ Phys.\          \vol(#1,#2,#3)}
\def \PL(#1,#2,#3){Phys.\ Lett.\          \vol(#1,#2,#3)}
\def \PRL(#1,#2,#3){Phys.\ Rev.\ Lett.\   \vol(#1,#2,#3)}
\def \PRp(#1,#2,#3){Phys.\ Rep.\          \vol(#1,#2,#3)}
\def \PR(#1,#2,#3){Phys.\ Rev.\           \vol(#1,#2,#3)}
\def \PTP(#1,#2,#3){Prog.\ Theor.\ Phys.\ \vol(#1,#2,#3)}
\def \ibid(#1,#2,#3){{\it ibid.}\         \vol(#1,#2,#3)}

\def\thebibliography#1{\par\newpage
\section*{References\@mkboth
  {REFERENCES}{REFERENCES}}\list
  {[\arabic{enumi}]}{\setlength\labelwidth{2ex}
   \setlength\labelsep{0.05in}%
   \setlength\leftmargin{0.25in}
   \setlength\itemsep{0pt}
    \usecounter{enumi}}
    \def\newblock{\hskip .11em plus .33em minus -.22em}
    \sloppy
    \sfcode`\.=1000\relax}

\def\@bibitem#1{\item\if@filesw \immediate\write\@auxout
       {\string\bibcite{#1}{\the\c@enumi}}\fi\ignorespaces
       {\ifnoncomplete\reversemarginpar{\hspace*{-1.05in}\makebox[1in][l]
       {{\footnotesize{\sl [#1]}}}}\fi}%
       }

\def\@cite#1#2{\unskip\nobreak\relax
    {[#1]}} 

\def\citenum#1{{\def\@cite##1##2{##1}\cite{#1}}}
\def\citea#1{\@cite{#1}{}}

\newcount\@tempcntc
\def\@citex[#1]#2{\if@filesw\immediate\write\@auxout{\string\citation{#2}}\fi
  \@tempcnta\z@\@tempcntb\m@ne\def\@citea{}\@cite{\@for\@citeb:=#2\do
    {\@ifundefined
       {b@\@citeb}{\@citeo\@tempcntb\m@ne\@citea\def\@citea{,}{\bf ?}\@warning
       {Citation `\@citeb' on page \thepage \space undefined}}%
    {\setbox\z@\hbox{\global\@tempcntc0\csname b@\@citeb\endcsname\relax}%
     \ifnum\@tempcntc=\z@ \@citeo\@tempcntb\m@ne
       \@citea\def\@citea{,}\hbox{\csname b@\@citeb\endcsname}%
     \else
      \advance\@tempcntb\@ne
      \ifnum\@tempcntb=\@tempcntc
      \else\advance\@tempcntb\m@ne\@citeo
      \@tempcnta\@tempcntc\@tempcntb\@tempcntc\fi\fi}}\@citeo}{#1}}
\def\@citeo{\ifnum\@tempcnta>\@tempcntb\else\@citea\def\@citea{,}%
  \ifnum\@tempcnta=\@tempcntb\the\@tempcnta\else
   {\advance\@tempcnta\@ne\ifnum\@tempcnta=\@tempcntb \else \def\@citea{--}\fi
    \advance\@tempcnta\m@ne\the\@tempcnta\@citea\the\@tempcntb}\fi\fi}

\def\affiliation#1{\cr
\makebox[0in]{\parbox{8in}{\begin{center} {\sl #1}\end{center}}} \cr}
\def\@affiliation{}

\def\and{\cr \makebox[0in]{\rule[-1cm]{0mm}{1cm}and } \cr}

\def\maketitle{\par
 \begingroup
 \def\thefootnote{\fnsymbol{footnote}}
 \def\@makefnmark{\hbox
 to 0pt{$^{\@thefnmark}$\hss}}
 \if@twocolumn
 \twocolumn[\@maketitle]
 \else \newpage
 \global\@topnum\z@ \@maketitle \fi\thispagestyle{plain}\@thanks
 \endgroup
 \setcounter{footnote}{0}
 \let\maketitle\relax
 \let\@maketitle\relax
 \gdef\@thanks{}\gdef\@author{}\gdef\@title{}
 \gdef\@affiliation{} \let\affiliation\relax	%
 \let\thanks\relax}

\def\@maketitle{\newpage
 \null
 \vskip 0em plus 2em minus 0em     
 \ifx\@date\@empty\else
   \begin{flushright}
    {\normalsize \@date}\\         
    {\ifnoncomplete(\today)\fi}      
   \end{flushright}
   \vskip 3em plus 2em minus 2em   
 \fi
 \begin{center}
  {\frtnsfb \@title \par}     
  \vskip 3em plus 1em minus 1.5em  
  {
   \lineskip .5em plus 0em minus .3em   
   \begin{tabular}[t]{c}\@author
   \end{tabular}\par}
\end{center}
 \par
 \vskip 6em plus 2em minus 4em}     

\def\abstract{\if@twocolumn
\section*{Abstract}
\else \normalsize
\fi}

\def\endabstract{\if@twocolumn\fi\par\clearpage}

\def\section{\@startsection {section}{1}{\z@}{3.5ex plus 1ex minus
    .2ex}{2.3ex plus .2ex}{\normalsize\bf}}
\def\subsection#1{\subsectioncom{\sc{#1}}}
\def\subsectioncom{\@startsection{subsection}{2}{\z@}
    {3.25ex plus 1ex minus .2ex}{1.5ex plus .2ex}{\small}}
\def\subsubsection{\@startsection{subsubsection}{3}{\z@}{3.25ex plus
1ex minus .2ex}{1.5ex plus .2ex}{\small}}

\def\@addmarginpar{\@next\@marbox\@currlist{\@cons\@freelist\@marbox
    \@cons\@freelist\@currbox}\@latexbug\@tempcnta\@ne
    \if@twocolumn
        \if@firstcolumn \@tempcnta\m@ne \fi
    \else
      \if@mparswitch
         \ifodd\c@page \else\@tempcnta\m@ne \fi
      \fi
      \if@reversemargin \@tempcnta -\@tempcnta \fi
    \fi
    \ifnum\@tempcnta <\z@  \global\setbox\@marbox\box\@currbox \fi
    \@tempdima\@mparbottom \advance\@tempdima -\@pageht
       \advance\@tempdima\ht\@marbox \ifdim\@tempdima >\z@
      \else\@tempdima\z@ \fi
    \global\@mparbottom\@pageht \global\advance\@mparbottom\@tempdima
       \global\advance\@mparbottom\dp\@marbox
       \global\advance\@mparbottom\marginparpush
    \advance\@tempdima -\ht\@marbox
    \global\ht\@marbox\z@ \global\dp\@marbox\z@
    \vskip -\@pagedp \vskip\@tempdima\nointerlineskip
    \hbox to\columnwidth
      {\ifnum \@tempcnta >\z@
          \hskip\columnwidth \hskip\marginparsep
        \else \hskip -\marginparsep \hskip -\marginparwidth \fi
       \box\@marbox \hss}
    \vskip -\@tempdima
    \nointerlineskip
    \hbox{\vrule \@height\z@ \@width\z@ \@depth\@pagedp}}

\def\ref#1{
    \@ifundefined{r@#1}{{#1}\@warning{Reference `#1'
    on page \thepage \space
    undefined}}{\edef\@tempa{\@nameuse{r@#1}}\expandafter
    \@car\@tempa \@nil\null}}
\def\refn#1{\@ifundefined{r@#1}{{#1}\@warning{Reference `#1'
    on page \thepage \space
    undefined}}{\edef\@tempa{\@nameuse{r@#1}}\expandafter
    \@car\@tempa \@nil\null}}

\def\endequationl{\eqno \@eqnnum 
$$\global\@ignoretrue}

\def\eqnarray{\stepcounter{equation}\let\@currentlabel=\theequation
\global\@eqnswtrue
\global\@eqcnt\z@\tabskip\@centering\let\\=\@eqncr
$$\arraycolsep\z@
\halign to \displaywidth\bgroup\@eqnsel\hskip\@centering
  $\displaystyle\tabskip\z@{##}$&\global\@eqcnt\@ne
  \hskip 2\arraycolsep \hfil$\displaystyle{{}##{}}$\hfil
  &\global\@eqcnt\tw@ \hskip 2\arraycolsep
  $\displaystyle\tabskip\z@{##}$\hfil
   \tabskip\@centering&\llap{##}\tabskip\z@\cr}
\def\mmodetrue{\mmode=\iftrue}

\def\eqnarrayl#1{\stepcounter{equation}\let\@currentlabel=\theequation
\label {#1}
\global\@eqnswtrue
\global\@eqcnt\z@\tabskip\@centering\let\\=\@eqncr
$$\arraycolsep\z@
\halign to \displaywidth\bgroup\@eqnsel\hskip\@centering
  $\displaystyle\tabskip\z@{##}$&\global\@eqcnt\@ne
  \hskip 2\arraycolsep \hfil$\displaystyle{{}##{}}$\hfil
  &\global\@eqcnt\tw@ \hskip 2\arraycolsep
  $\displaystyle\tabskip\z@{##}$\hfil
   \tabskip\@centering&\llap{##}\tabskip\z@\cr}
\def\label#1{
\@bsphack\if@filesw {
{\ifnoncomplete{\makebox[1in][r]{\footnotesize{\sl [#1]}}}\fi}%
\let\thepage\relax
   \xdef\@gtempa{\write\@auxout{\string
      \newlabel{#1}{{\@currentlabel}{\thepage}}}}
}\@gtempa
   \if@nobreak \ifvmode\nobreak\fi\fi\fi\@esphack}

\def\newlabel#1#2{
\@ifundefined{r@#1}{}{\@warning{Label `#1' multiply
   defined}}\global\@namedef{r@#1}{#2}}

\def\endeqnarrayl{\@@eqncr\egroup
      \global\advance\c@equation\m@ne$$\global\@ignoretrue}

\newif\if@numbersec \@numbersectrue
\def\appendix{\par\clearpage
  \setcounter{section}{0}
  \setcounter{subsection}{0}
  \def\thesection{\Alph{section}}
  \def\thesubsection{\arabic{subsection}}
  \@ifstar{\def\@sectname{Appendix}\@numbersecfalse}
          {\def\@sectname{Appendix~}\@numbersectrue}}

\def\thefigures#1{\par\clearpage\section*{Figures\@mkboth
  {FIGURES}{FIGURES}}\list
  {Fig.~\arabic{enumi}.}{\labelwidth\parindent\advance\labelwidth -\labelsep
      \leftmargin\parindent\usecounter{enumi}}}
\def\figitem#1{\item\label{fig:#1}}

\def\thetables#1{\par\clearpage\section*{Tables\@mkboth
  {TABLES}{TABLES}}\list
  {Table~\arabic{enumi}.}{\labelwidth-\labelsep
      \leftmargin0pt\usecounter{enumi}}}

\def\@sect#1#2#3#4#5#6[#7]#8{\ifnum #2>\c@secnumdepth
     \def\@svsec{}\else
     \refstepcounter{#1}\edef\@svsec{\ifnum #2=1 \@sectname
         \if@numbersec\csname the#1\endcsname\fi.\else
         \csname the#1\endcsname.\fi
        \hskip 1em }\fi
     \@tempskipa #5\relax
      \ifdim \@tempskipa>\z@
        \begingroup #6\relax
          \@hangfrom{\hskip #3\relax\@svsec}{\interlinepenalty \@M #8\par}
        \endgroup
       \csname #1mark\endcsname{#7}\addcontentsline
         {toc}{#1}{\ifnum #2>\c@secnumdepth \else
                      \protect\numberline{\csname the#1\endcsname}\fi
                    #7}\else
        \def\@svsechd{#6\hskip #3\@svsec #8\csname #1mark\endcsname
                      {#7}\addcontentsline
                           {toc}{#1}{\ifnum #2>\c@secnumdepth \else
                             \protect\numberline{\csname the#1\endcsname}\fi
                       #7}}\fi
     \@xsect{#5}}

\def\@sectname{}

\def\Figure #1{Fig.~#1{\ifnoncomplete\marginpar{\footnotesize\sl fig [#1]}\fi}}
\def\Figures #1{Figs.~#1{\ifnoncomplete\marginpar{\footnotesize\sl fig
[#1]}\fi}}

\def \@magscale#1{ scaled \magstep #1}
\font\frtnsfb = cmssbx10 \@magscale2 
\newcount\ieq
\newcount\jeq
\newcount\keq
\def \eq{
\multiply\ieq by 2
\jeq=\ieq
\divide\jeq by 4
\multiply\jeq by 4
\ifnum\ieq=\jeq \end{eqnarray} \keq=1 
\else
\keq=2 \begin{eqnarray} \fi
\ieq=\keq
}
\let \eqend=\eq
\def \mathbox(#1){\invisible\ifmmode{{#1}}\else{\mbox{${#1}$}}\fi}
\def \mbf(#1){\mbox{\boldmath{$#1$}}}
\def \abs(#1){\mathbox(\left|{#1}\right|)}
\def \bracket(#1){\mathbox(\left\langle{#1}\right\rangle)}
\def \brav(#1){\mathbox(\langle {#1}|)}
\def \cg(#1,#2,#3,#4,#5,#6){\mathbox({(#1\,#2\,#3\,#4|#5\,#6)})}
\def \comm(#1,#2){\mathbox(\left[{#1},{#2}\right])}
\def \dfdx(#1,#2){\mathbox(\frac{{\rm d}{#1}}{{\rm d}{#2}})}
\def \delfdelx(#1,#2){\mathbox(\frac{\partial{#1}}{\partial{#2}})}
\def \inprod(#1,#2){\mathbox({(#1\cdot #2)})}
\def \inprodij(#1){\mathbox({\inprod(#1_i,#1_j)})}
\def \intd(#1,#2){\mathbox({\int^#1_#2 \; \rmd})}
\def \eps(#1){\mathbox(\epsilon_{#1})}
\def \half(#1){\mathbox(\frac{#1}{2})}
\def \ketv(#1){\mathbox(|{#1}\rangle)}
\def \matele(#1,#2,#3){\mathbox(\left\langle {#1}|\,{#2}\,|{#3}\right\rangle)}
\def \mateled(#1,#2,#3){\mathbox(\left\langle
{#1}||\,{#2}\,||{#3}\right\rangle)}
\def \hatmbf(#1){\mathbox({\hat{\mbf({#1})}})}
\def \ninej(#1,#2,#3,#4,#5,#6,#7,#8,#9){\mathbox(\left\{\matrix
     {#1&#2&#3\cr#4&#5&#6\cr#7&#8&#9\cr}\right\})}
\def \rtov(#1,#2){\mathbox(\sqrt{{#1\over #2}})}
\def \sixj(#1,#2,#3,#4,#5,#6){\mathbox(\left\{\matrix
     {#1&#2&#3\cr#4&#5&#6\cr}\right\})}
\def \third(#1){\mathbox(\frac{#1}{3})}
\def \Trace(#1){\mathbox({\hbox{Tr} \left\{#1\right\}})}
\def \outprod(#1,#2){\mathbox({(#1\times #2)})}

\def \abs#1{\left|{#1}\right|}
\def \bra{\mathbox(\langle)}

\def \ie{{\it i.e.}}
\def \invisible{\mbox{$\rule{0mm}{1mm}$}}
\def \ket{\mbox{$\rangle$}}

\def \lam{\mbox{$\lambda$}}
\def \Nc{\mbox{$N_c$}}

\def \rmd{{\rm d}}

\endinput

\\ sty2.sty

\def\@cite#1#2{\unskip\nobreak\relax
                   {[#1]}} 
\def\section{\@startsection {section}{1}{\z@}{3.5ex plus 1ex minus
    .2ex}{2.3ex plus .2ex}{\normalsize\bf}}
\def\subsection#1{\subsectioncom{{\it #1}}}

\endinput

\documentstyle[12pt,sty1,sty2]{article}

\def\und#1{{\bf #1}}
\def \pnum #1,#2,#3.{\und{#1} (19#3) #2}
\def \PRD  #1,#2,#3.{Phys.\ Rev.\ D \pnum {#1},{#2},{#3}.}
\def \PRC  #1,#2,#3.{Phys.\ Rev.\ C \pnum {#1},{#2},{#3}.}
\def \PRA  #1,#2,#3.{Phys.\ Rev.\ A \pnum {#1},{#2},{#3}.}
\def \PRL  #1,#2,#3.{Phys.\ Rev.\ Lett.\ \pnum {#1},{#2},{#3}.}
\def \PL   #1,#2,#3.{Phys.\ Lett.\ \und{B} \pnum {#1},{#2},{#3}.}

\def \NP   #1,#2,#3.{Nucl.\ Phys.\ \pnum {#1},{#2},{#3}.}
\def \Prog #1,#2,#3.{Prog.\ Theor.\ Phys.\ \pnum {#1},{#2},{#3}.}
\let\PTP=\Prog
\def \PRep #1,#2,#3.{Phys.\ Rep.\ \pnum {#1},{#2},{#3}.}
\def \Zeit #1,#2,#3.{Z.\ Phys.\ \pnum {#1},{#2},{#3}.}

\def\ibid #1,#2,#3.{{\it ibid.}\ \pnum #1,#2,#3.}
\def \prepr #1,#2.{\def\secondpara{#2}
                 {#1} preprint\ifx\secondpara\empty\else{ ({#2})}\fi}

\def\MO{M.\ Oka}   
\def\AM{R.D.\ Amado}

\def\ie{{\it i.e.}}

\def\wave{\sim}
\def\abs#1{\left|{#1}\right|}
\def\ket #1>{|{#1}\rangle} \def\bra #1|{\langle {#1}|}
\def\braket <#1>{\langle {#1}\rangle}
\def\mate <#1|#2|#3>{\bra {#1}|\,{#2}\,\ket {#3}>}
\def\mated <#1||#2||#3>{\bra {#1}||\,{#2}\,|\ket {#3}>}
\def\Tr{\hbox{Tr}}
\def\Trace <#1>{\Tr \left\{#1\right\}}
\def\comm[#1,#2]{\left[{#1},{#2}\right]}
\def\dotp(#1.#2){#1\cdot #2}
\def\xp(#1.#2){{#1\times #2}}

\def\diag{\mathop{\rm diag}\nolimits}

\def\cg(#1,#2,#3,#4,#5,#6){(#1,#2,#3,#4|#5,#6)}
\def\sixj(#1,#2,#3,#4,#5,#6)
    {\left\{\matrix {#1&#2&#3\cr#4&#5&#6\cr}\right\}}

\def\Tr{\mathop {Tr}}
\def\nab{\vec\nabla}
\def\lam{\lambda}
\def\tr{\mathop {\rm tr}}
\def\Fpi{F_{\pi}} \def\fpi{\Fpi}
\def\mpi{m_{\pi}}

\def\xv{\vec x}
\def\Rv{\vec R}
\def\ie{{\it i.e.\relax}}
\def\diag{\mathop {\rm diag}}
\def\wave{\sim}

\def\bra{\langle}
\def\ket{\rangle}

\def\Nc{N_c}
\def\sv{\vec\sigma} \def\tv{\vec\tau}

\begin{document}

\final

\date{TIT/TP-209/NP\\
      January, 1993}

\title{Nuclear Force in the Skyrme Model: Quasistatic Approach}

\author{Makoto Oka\\
Department of Physics,
Tokyo Institute of Technology\\
Meguro, Tokyo 152, Japan\\}

\maketitle

\abstract {
Dynamics of two-Skyrmion systems is studied in the quasistatic
approach.
The quasistatic approach enables us to formulate the collective
coordinate quantization of two-Skyrmion systems consistently with the
$1/\Nc$ expansion.
By constructing quasistatic field configurations for largely separated
two Skyrmions,
their energies are computed for three independent relative
orientations of the Skyrmions in the isospin space.
The obtained energies are much lower than those for the product
approximation.
In particular, the repulsion in
the two-pion range is considerably reduced.\\}

\thispagestyle{empty}

\newpage

\section{Introduction}
   The Skyrme model{\cite{Sky,ANW}} of the nucleon provides us with a
completely new
picture of the nucleon and yet a somewhat old aspect of the roles of
the pion in the nucleus.  The Skyrmion is a static pion field
configuration, which describes a QCD baryon.
The baryon number, not associated with the quark, is identified with the
winding number, which characterizes
a nontrivial topology of the pion field configuration.
The model successfully explains various properties of the nucleon,
including the electromagnetic and weak verteces, and the couplings to the
pion and the other mesons.
One of the advantages of the Skyrme model is that it
takes account of the chiral symmetry of QCD consistently in the
baryonic sector.
It is well known that the effective chiral theory represents the dynamics
of the low lying mesons very well and predicts various useful relations.
Similarly, the Skyrme model gives a chiral symmetric description of the
nucleon structure and the meson-nucleon dynamics{\cite{Zahed}}.

When we apply the Skyrme model to the nuclear force problem,
we follow the conventional approach, that is, an adiabatic approach,
where the static potential is computed first and then
nonstatic effects are taken into account later{\cite{Jackson,VM,Oka,OH}}.
This approach is quite successful in computing the contributions of
meson exchanges to the nuclear force.
In the Skyrme model, we consider two steps:
first, find an appropriate two-Skyrmion configuration for
each Skyrmion-Skyrmion separation.
The static energy of a set of such configurations for various
separations gives an adiabatic inter-Skyrmion potential.
Then the nucleon-nucleon interactions are computed from the
Skyrmion-Skyrmion interactions.
It happens that each of these two steps has some technical difficulties.
For the second step, one has to project the Skyrmion spin and isospin
into the nucleon.  At large separations, each Skyrmion can be rotated
and is projected individually.  At shorter distances, however, the
Skyrmion may lose its identity and thus we cannot define individual
rotations.
We will not address this problem further in this article.  Instead, in
sect.~4, we use an approximation which is valid at large separations.

The first step also involves a difficulty.
Skyrme proposed to employ the product approximation (PA) for the two-Skyrmion
configurations and showed that the model
gives the correct one-pion exchange (OPE) potential
between two nucleons.
PA, however, fails to explain the  medium range attraction, which is
needed for the nuclear binding.
Various alternatives have been proposed.
They suggest that the strong medium-range repulsion
in the product approximation is not real.
Ref.\cite{Wal}, for instance, the authors derive a mild attraction at
medium distances using a field configuration numerically found by
minimizing the static energy at each separation.
A semi-analytic method proposed in ref.\cite{AM,AM-hosaka} gives a
weak repulsion
at medium distances.

Nonstatic effects on the adiabatic configurations are regarded as
correctioons of higher order in $\hbar/\Nc$ and require quantization
of the system with the static configuration as a classical background
field of order $\sqrt{\Nc}$.
A systematic $1\over\Nc$ expansion
is essential for a consistent quantization of two-skyrmion systems.
The aim of this paper is to present a study of
the Skyrmion-Skyrmion interaction using a novel quasistatic
equation for the two-Skyrmion system,
which provides us with the classical field configuration consistent
with a quantization in the collective coordinate approach.
(A preliminary result has been reported in ref.\cite{PRL}.)
The Skyrmion-Skyrmion distance is taken as a collective coordinate.
This is an approach most suitable at large separation because there the
relative motion is slow and is to be treated adiabatically.
The quantization for excitation modes which are orthogonal
to the collective motion is treated separately from the relative motion.

In the present paper, we study the quasistatic equation and its
solutions.  The adiabatic potential energy is calculated and is
compared with the product approximation.  We also study the
semiclassical quantization of the SU(2) rotations of the Skyrmion in
order to get the nucleon-nucleon interaction.
In sect.~2, we present the basic formulation of the quasistatic approach
as well as the product approximation.  The quasistatic equation is
solved in sect.~3 and the nuclear force is calculated in sect.~4.  A
conclusion is given in sect.~5.

\section{Formulation}
\subsection{Skyrme Lagrangian}

\def\vphi{\vec\phi}
      We choose the original Skyrme model
with  a finite pion mass term.
(It should be noted, however, that our formulation is independent of the
particular choice of the Lagrangian.)
\par
\eq
     {\cal L} &=& {\Fpi^2\over 16} \tr [\partial_{\mu}U \partial^{\mu}
         U^{\dagger} ] +{1\over 32 e^2} \tr[ \partial_{\mu}UU^{\dagger},
          \partial_{\nu}U U^{\dagger}]^2
            + {\mpi^2\fpi^2\over 8} \tr[U-1] \cr
     &=& {1\over 8} (\partial_{\mu}\phi_k)^2 -{1\over 4} \left[
       (\partial_{\mu}\phi_k)^2 (\partial_{\nu}\phi_{\ell})^2 -
       (\partial_{\mu}\phi_k \partial^{\mu}\phi_{\ell})^2 \right]
              +{\mu^2\over 4} (\phi_4 -1)
\label{eq:Lag}
\eq
where in the second line we define $U=\phi_4 + i\tv\cdot\vphi$ an
employ a dimensionless unit (Skyrmion
unit, SU) with length measured in $(\Fpi e)^{-1}$ and energy in $\fpi/e$.
$\mu$ is the inverse Compton wave length of the pion in SU, $\mu =
m_{\pi}/\fpi e$.
The four component field $\phi_k$ ($k=1,\ldots 4$)
should satisfy
the unitarity normalization
condition, $\sum_{k=1}^4 \phi_k^2 = 1$,
which can be imposed by adding the Lagrange multiplier term
\eq
        {\cal L} \to {\cal L} + {\lambda\over 2} (\phi_k^2 -1)
\eq
The Euler-Lagrange equation is then given by
\eq
      \partial_{\mu} ({\cal M}_{k\ell}\partial^{\mu}\phi_{\ell}) -
          \delta_{k4} {\mu^2\over 4} -   \lam\phi_k = 0
\eq
where
\eq {\cal M}_{k\ell} [\phi] \equiv {1\over 4}\delta_{k\ell}
             - (\partial_{\mu} \phi_m)^2\delta_{k\ell}
             + \partial_{\mu} \phi_k \partial^{\mu} \phi_{\ell}
\eq
For a static configuration, the energy is given by
\eq
      E[\phi] = \int \left\{ {1\over 8} \, (\nab \phi_k)^2
                +{1\over 4} \,[ (\nab \phi_k)^2  (\nab\phi_{\ell})^2
                - (\nab \phi_k \cdot \nab \phi_{\ell})^2]
                      +{\mu^2\over 4}  (1-\phi_4) \right\}\, d^3x
\label{eq:se}
\eq

\subsection{Product Approximation (PA)}
  In constructing an adiabatic Skyrmion-Skyrmion potential, one needs
a general static field configuration describing two Skyrmions separated
by a distance $R$.
A simple prescription for this was given by Skyrme \cite{Sky}.  He
showed that a
product of two unitary matrix fields with baryon numbers $B_1$ and
$B_2$, respectively, yields a new field configuration with total baryon
number $B=B_1 + B_2$.  Thus by choosing $B_1=B_2=1$, one can construct a
two-baryon configuration (product approximation, PA) from the $B=1$ Skyrmion
solution:
\eq
     U_{PA} = U_1\cdot U_2 = A U_h(\xv-{\Rv\over 2}) A^{\dagger}
                             B U_h(\xv+{\Rv\over 2}) B^{\dagger}
\label{eq:PA}
\eqend
where $U_h$ is a hedgehog,
\eq
      U_h(\xv) = \exp [ i\vec\tau\cdot\hat x f(x)]
\label{eq:hhg}
\eqend
and $A$ and $B$ are constant $SU(2)$ matrices describing rotations of
each Skyrmion.
This configuration is useful at large separation, giving a
correct one-pion exchange (OPE) potential, while it has a serious
defect at shorter separation, that is, it breaks a reflection symmetry.
This problem is caused by the asymmetry,
$U_1\cdot U_2 \neq U_2\cdot U_1$, and
can be seen for $A=B=1$,
from
\eq
   U_1\cdot U_2 - U_2\cdot U_1 = 2i \vec\tau\cdot (\xv\times\Rv)
              {1\over r_1 r_2} f(r_1) f(r_2)
\label{eq:diff}
\eqend
where $r_{1,2} \equiv \abs{\xv\mp \Rv/2}$.
Eq.(\ref{eq:diff}) shows that the difference of $U_1\cdot U_2$ and
$U_2\cdot U_1$
looks like a ``magnetic field'' around a current in $\Rv$ direction
and therefore differs in symmetry from their common part, which is a
rotation-free (electric) field.
In particular, the $U_{PA} = \phi_4 + i\vec\tau\cdot\vec\phi$
configuration breaks the reflection symmetry,
\eq
     \phi_x(-x,y,z) = -\phi_x (x,y,z)
\eqend
which is satisfied by a single hedgehog and therefore an
infinitely-separated two hedgehogs.
Thus we conclude $U_{PA}$ contains an extra ``iso-magnetic'' component
and does not give the most appropriate configuration for the adiabatic
potential.

\subsection{Quasistatic Configuration (QS)}
It is not possible in general to find a static two-Skyrmion solution
of the equation of motion because
the Skyrmions interact with each other and therefore do not form a
static configuration.
One only finds either time-dependent (scattering) solutions or special
solutions, such as
(A) two Skyrmions in the limit of large separation $R\to\infty$,
(B) the $B=2$ hedgehog solution, or
(C) a bound solution with axial (toroidal) symmetry,
which is believed to have the minimum energy \cite{donuts}.
The energy of the $B=2$ hedgehog (B) lies about 2 GeV above
the threshold, \ie, the energy of two isolated Skyrmions (A) ($=2M_0$,
$M_0$ being the single Skyrmion energy).
The energy of the bound state (C) is about 70 MeV below that
threshold.
These solutions, however, are not sufficient to describe
the general two-Skyrmion dynamics, especially,
at medium and large separations \cite{Manton}.

  We here consider a so-called quasistatic configuration instead.
In our previous papers \cite{paperI} (hereafter referred as I),
we have proposed to describe two Skyrmions at large separation
by a field configuration given as a solution
of a quasistatic equation.
The formalism  is based on the collective coordinate method,
which requires the following conditions:
(1) Kinetic energies for both the translational and
rotational motions of the individual Skyrmions are small.
This condition is satisfied for the large $\Nc$ theory, because
the kinetic energy is of higher order in $1/\Nc$.
(2) The intersoliton interaction is weak.  This is satisfied
for large separation $R$ and the interaction is
dominated there by the one-pion exchange (OPE)
potential proportional to $\epsilon\equiv\exp(-\mu R)/R$.
By making a double expansion in $1/\Nc$ and $\epsilon$,
we derive a quasistatic equation
(QSE) for a static
$B=2$ configuration with fixed $R$.
Although the configuration
is not stable  because of the intersoliton interaction,
it can be quantized by applying the Dirac
quantization method with constraints.
\par

\def\vzeta{\vec\zeta}
\def\vnab{\vec\nabla} \def\avec{\vec a}
The static equation of motion does not yield a general two-Skyrmion
solution.   We introduce an extra source term corresponding to a force
field which holds two Skyrmions at a given distance.
Then the quasistatic (QS) equation is given by
\eq
      \vnab\cdot\left( M_{k\ell}\vnab\phi_{\ell}\right) + \delta_{k4}
         {\mu^2\over 4} +    \lam\phi_k =
        M_{k\ell} \avec\cdot\vzeta_{\ell}
\label{eq:QSE}
\eq
where
\eq M_{k\ell} [\phi] \equiv {1\over 4}\delta_{k\ell}
             + (\nab \phi_m)^2\delta_{k\ell}
             -  \nab \phi_k\cdot\nab \phi_{\ell}
\eq
The force term (the right hand side of eq.(\ref{eq:QSE}) is
proportional to the classical acceleration,
\eq
      \avec(r)\equiv \ddot{\vec R}  = - {2\over M_0} \vnab V(r)
\label{eq:acc}
\eq
and the low-energy mode corresponding to the relative motion of the
Skyrmions,
\eq
     \vzeta_{\ell} = \vnab_R \phi_{\ell}
\eq
This is the mode which describes fluctuation of the distance of two
Skyrmions and therefore its energy is of order $\epsilon$ and
vanishes when $R\to \infty$.

The physical meaning of the force term can be understood in several
ways.
First, it can be
obtained from the full time-dependent equation of motion,
assuming that $\phi$ depends on time only through the collective
coordinate $R$ and then setting $\dot R=0$ but allowing
non-zero $\ddot R$.
The solution is then
the solution of the full equation at a turning point
where $\dot R$ is zero but $\ddot R$ is not.
In this sense, this solution may be called ``quasistatic'' (QS).
\par

Another derivation of the QS equation (\ref{eq:QSE}) is more appropriate
in the context of quantization of the system.
It is easy to show that the equation
is given by the variational principle for the static energy (\ref{eq:se}),
\eq
      {\delta E(R)\over \delta\phi_k} = 0
\eq
under a condition
\eq
     (\delta\phi\cdot\vzeta) \equiv
    \int \delta \phi_k \, M_{k\ell}\, \vzeta_{\ell} = 0
\label{eq:cond}
\eq
This variation makes the solution of QSE minimize the static energy
with respect to variations that do not change the separation of two
Skyrmions.
This property of the QS solution is crucial when it comes to
quantize the system by the collective coordinate method.
The quantum description of the system concerns two sets of variables,
collective and noncollective.
The collective variable is the one which is associated with
a zero (ZM) or nearly-zero (NZM) energy mode of field fluctuation.
In the two-Skyrmion system,
there are six ZM's, three center of mass translational
motions and three overall rotations, and
six NZM's, three relative translational motions and
three relative orientations.
These NZM's have energy of order $\epsilon$ and cause
the instability of the static configuration.
Quantization of the noncollective field
fluctuations can be done
under constraints which exclude ZM and NZM's.
The background (classical) configuration for the quantization
must be stable against noncollective
field fluctuations.

\def\phiQ{\phi^{QSE}} \def\dphi{\delta\phi}

One can see the role of the QSE in the quantization by the expansion
of the Lagrangian density with respect to field fluctuations around
the QSE solution.
For $\dphi_k = \phi_k - \phiQ_k$,
\eq
   {\cal L}[\phi_k] = {\cal L} [\phiQ_k] +
        {\partial{\cal L}\over \partial \phiQ_k} \,\dphi_k
    + {1\over 2} \,{\partial^2{\cal L}\over
         \partial\phiQ_k\partial\phiQ_{\ell}}\, \dphi_k\dphi_{\ell}
    + \cdots
\label{eq:Lexp}
\eq
In the case of the single Skyrmion, $\phi_k ({\rm Skyrmion})$
satisfies the static equation motion and therefore the second term of
the expansion vanishes.  That is, however, not possible for general two
Skyrmion systems.
The QSE solution gives instead
\eq
        {\partial{\cal L}\over \partial \phiQ_k} =  \left(M_{k\ell}
            \avec\cdot\vzeta_{\ell} \right)^{\dagger}
\eq
The nonvanishing linear term in (\ref{eq:Lexp}) leads us to an
instability of the system if we consider general
$\dphi_k$.
In the quantum description, $\dphi_k$ is quantized as a pion.
The linear term describes a
Yukawa coupling of the pion to the Skyrmion of order
$\sqrt{\Nc}$.  Then because exchanges of pions yield energy
of order $\Nc$, $E_{QS} = -{\cal L}[\phiQ]$ does not give the full
$O(\Nc)$ classical energy of the system.

We can remedy this problem by introducing the collective coordinate
$\vec R$.  The NZM, $\vzeta$, is then taken care of the equation for
$\vec R$.
In order to avoid the redundancy of variables as a result of
introducing the collective variable, one has to impose an
orthogonality of general field fluctuation modes to the collective motion.
The necessary condition happens to be eq.(\ref{eq:cond}),
which makes the noncollective modes orthogonal to the $\vec R$ motion.
Also itguarantees that the integral of the
second term of eq(\ref{eq:Lexp}) vanishes and  so does the
linear coupling of the non-collective pion modes to the Skyrmion of
order $O(\sqrt{\Nc})$.
Now solutions of the QSE are stable against noncollective pion field
fluctuations  and yield a manifold as a family of saddle
points for each $R$.
The manifold for a given $R$ will differ by a small amount of order
$\epsilon$ from that for another $R$.
\par

Technical problems in the quantization under the condition (\ref{eq:cond})
are known to be overcome by using the Dirac quantization method for
constrained systems\cite{Dirac}.
In paper (I),
we have shown that Dirac quantization
can be consistently performed
with solutions of the QSE as the background configuration.
The condition (\ref{eq:cond}) eliminates spurious leading order
$\pi$--Skyrmion coupling and therefore allows a clear separation
of the classical and quantal contributions.
Solutions of the QSE and their usefulness were demonstrated in (I)
for simple ($1+1$)-dimensional soliton models.
In this paper, we apply the same formalism to the Skyrme model,
a realistic 3-dimensional problem.

\subsection{Linearization of Quasistatic Equation}

\def\phiz{\phi^0}
In order to solve the QS equation, we employ a long-distance
expansion, \ie, an expansion with respect to
$\epsilon\equiv{e^{-\mu R}\over \mu R}$.
We take the reflection symmetric part of
the product (eq. (\ref{eq:PA})) of two Skyrmion fields
as a trial solution, $\phiz_k$.
Then assuming
$\phi_k-\phiz_k$ is of order $\epsilon$, we expand
each of the following quantities in  $\epsilon$,
\eq
     \phi_k&=&\phiz_k +\epsilon\varphi_k \cr
     \vzeta_k &=& \vzeta^0_k + \epsilon\vzeta^1_k
               \equiv  \vnab_R\phiz_k  + \epsilon\vzeta^1_k  \cr
     \lambda &=& \lambda_0 + \epsilon\lambda_1 \nonumber
\eq
Because the trial function satisfies the static equation of motion at
$R \to\infty$, one has
\eq
         W\equiv \epsilon w\equiv \vnab \cdot\left( M^0_{k\ell}\vnab
            \phiz_{\ell}\right)  + \delta_{k4} {\mu^2\over 4}
                  +\lambda_0\phiz_k \wave O(\epsilon)
\eq
Note that the asymmetric part, eq.(\ref{eq:diff}), vanishes at $R\to\infty$.
The adiabatic potential $V(R)$ is then
\eq
     V = E[\phiz_k] -2 M_0 + O(\epsilon^2) \wave O(\epsilon)
\eq
and the classical acceleration $\avec$ is similarly
\eq
     \avec \equiv \epsilon \avec_1 \wave V(R) \wave O(\epsilon)
\eq
Using the above expressions, one obtains, to the leading order in
$\epsilon$,
\eq
      \vnab\cdot\left( M^0_{k\ell}\vnab \varphi_{\ell}\right)
             +\vnab\left({\partial M_{k\ell}\over\partial\phiz_m}\,
               \vnab\phiz_{\ell} \,\varphi_m \right)
           +\lambda\varphi_k
              = - w - \lambda_1\phiz_k +M^0_{k\ell}
             \avec_1\cdot\vzeta^0_{\ell}
\label{eq:QSEL}
\eq
It should be noted that the solution of this linear equation for
$\varphi_k$ determines $\phi_k$ to the order $\epsilon$, while the
adiabatic potential $V[\phi_k]$ is given to $O(\epsilon^2)$, namely
\eq
       V[\phiz_k+\epsilon\varphi_k] = E[\phiz_k] - 2M_0 + \epsilon\,
{\partial E\over \partial \phiz_k}\, \varphi_k + \cdots
\eq
where ${\partial E\over \partial \phiz_k} \wave O(\epsilon)$ because of
the equation of motion for $\phiz_k$.
Once we solve this equation, the solution can be improved by
accounting the next order in $\epsilon$.

\section{Solutions of Quasistatic Equation}
We consider three series of solutions distinguished by their reflection
symmetries.
Each forms a path of field configurations
leading two separated Skyrmions to interacting ones
at shorter distances.
Obtained configurations give adiabatic potentials, which can be
compared with that for
those computed in the product approximation (\ref{eq:PA}).
These configurations for $m_{\pi}=0$ may also be obtained
approximately by using an
analytic method proposed by Atiyah and Manton\cite{AM}.
Adiabatic potentials for such cases are given in ref.\cite{AM-hosaka}.
\par

     The first set of configurations (labeled by H) satisfies the
reflection symmetry
of a system of two identical hedgehogs located at $z=\pm R/2$
($R \to\infty$), (see \Figure{1a})
\eq
    \phi_k (-x, y, z) &=& \diag (-1,1,1,1)_{k\ell}\,
                             \phi_{\ell} (x, y, z) \cr
    \phi_k (x, -y, z) &=& \diag (1,-1,1,1)_{k\ell}\,
                             \phi_{\ell} (x, y, z)
      \label{eq:Hsym} \\
    \phi_k (x, y, -z) &=& \diag (1,1,-1,1)_{k\ell}\,
                             \phi_{\ell} (x, y, z) \nonumber
\eq
Solutions of QSE with this symmetry will form a path leading
two isolated hedgehogs (A) into the $B=2$ hedgehog (B).
They can be compared with the product
approximation (PA) of two hedgehogs,
$U=U_1U_2$, \ie, $A=B=1$ in (\ref{eq:PA}).
Note that the asymmetric term of PA (\ref{eq:diff}) breaks the first
two symmetries in (\ref{eq:Hsym}).
This asymmetry is considered a deficiency of PA,
which does not follow the desired steepest descent path.
Our solutions with the symmetries (\ref{eq:Hsym}) for all (large) $r$ form
the correct path for two Skyrmion dynamics.
\par

The second path (labeled by X) connects the lowest energy configuration (C)
with two separated Skyrmions with relative rotation of
180$^\circ$ around the $x$ axis.
It has the following reflection symmetries: (\Figure{1b})
\eq
    \phi_k (-x, y, z) &=& \diag (-1,1,1,1)_{k\ell}\,
                             \phi_{\ell} (x, y, z) \cr
    \phi_k (x, -y, z) &=& \diag (1,-1,1,1)_{k\ell}\,
                             \phi_{\ell} (x, y, z)
    \label{eq:Xsym} \\
    \phi_k (x, y, -z) &=& \diag (1,-1,1,1)_{k\ell}\,
                             \phi_{\ell} (x, y, z) \nonumber
\eq
The product approximation with
\eq
     A^{\dagger}B = i\tau_x = \exp i{\tau_x\over 2}\pi
\eq
corresponds to this symmetry.  At small $R$, however, PA is not able
to describe the lowest energy axially symmetric $B=2$ configuration (C).

The third set of solutions (labeled by Z) is for two separated Skyrmions with
relative 180$^\circ$ around the $z$-axis: $A^{\dagger}B = i\tau_z$.
 (\Figure{1c})
\eq
    \phi_k (-x, y, z) &=& \diag (-1,1,1,1)_{k\ell}\,
                             \phi_{\ell} (x, y, z) \cr
    \phi_k (x, -y, z) &=& \diag (1,-1,1,1)_{k\ell}\,
                             \phi_{\ell} (x, y, z)
    \label{eq:Zsym} \\
    \phi_k (x, y, -z) &=& \diag (-1,-1,-1,1)_{k\ell}\,
                             \phi_{\ell} (x, y, z) \nonumber
\eq
It is known that this symmetry gives the strongest repulsive adiabatic
potential in the product approximation.

\par

We solve eq.(\ref{eq:QSEL}) by a relaxation method on a
three-dimensional lattice of size
$25\times 25\times 40$, $\Delta x=0.2$
for $x$, $y$, $z\ge 0$.
We need typically about 200 iterations for convergence.
After each iteration, $\lam_1$ is computed from the unitarity
condition
so that $\phi_k$ is always exactly normalized.
Numerical accuracy is checked by integrating the total baryon number
and the energy for PA on the same lattice.
We keep the errors less than 0.5\%.
We use the parameters $\fpi=108$ MeV, $e=4.84$ and $m_{\pi}=138$ MeV.

The results are summarized in \Figure{2}, where the calculated potential
energies are compared with the corresponding PA results.
One sees that the quasistatic configurations yield much lower energies
than the product approximations, especially in the H and X symmetries.
In fact, for the X symmetry, the product
approximation does not proceed toward the expected
the lowest energy $B=2$ bound solution, which
has the axial symmetry around the $x$-axis.
This is considered a serious defect of PA.

The reduction of the repulsion obtained in PA can be seen more clearly
by subtracting the one-pion exchange potential.  Note that the OPE
part is common to both PA abd QS configurations and is given
analytical.  One sees that the shorter-range components of the
adiabatic potential in \Figure{3} for the three orientations H, X and
Z.  Figure shows that the two-pion repulsion seen in PA is an artifact
of the approximation and, in fact, the two-pion exchange potential is
very weak in the Skyrme model.  This is one of our main conclusion in
the present study.
\par

\section{Nucleon-Nucleon Potential}

The hedgehog (\ref{eq:hhg}) cannot form an eigenstate of spin and
isospin because its pion field configuration has its isospin
proportional to $\hat r$.  In order to obtain spin-isospin
eigenstates, we need to
quantize the rotational motion of the Skyrmion, that is, to take
superpositions of different orientations.  For the two Skyrmion
system, the rotational motion of individual Skyrmion should be
considered.  One has to make a projection to appropriate spin and
isospin state in order to obtain the nucleon-nucleon potential from
the Skyrmion-Skyrmion potential.  It is conventionally performed in PA
by regarding the rotations $A$ and $B$ in eq.(\ref{eq:PA}) as
collective coordinates and quantizing them semiclassically\cite{Jackson,VM}.
The same method cannot be applied immediately to the quasistatic
configuration.  We, however, assume that the similar projection to the
$NN$ state is possible at least for a large $R$, that is, small
$\epsilon$. Under this assumption, one can parametrize the
Skyrmion-Skyrmion adiabatic potential by\cite{OBBA}
\eq
    V(R,A,B) = V_c + V_s W + V_t Z + \cdots
\label{eq:WZexp}
\eq
where
\eq
        W&=& T^{A}_{pi} T^{B}_{pi} \cr
        Z&=& T^{A}_{pi} T^{B}_{pj} \left[ 3\hat R_i \hat R_j
                      -\delta_{ij}\right]  \\
        T^{A}_{pi}&\equiv& {1\over 2} \tr[\tau_p A\tau_i A^{\dagger}] \nonumber
\eq
For the two nucleon system, $W$ and $Z$ are reduced to
\eq
     \matele(NN,W,NN)&=& {1\over 9} (\sv_1\cdot\sv_2)
           (\tv_1\cdot\tv_2)\cr
     \matele(NN,Z,NN)&=& {1\over 9} S_{12}^T
           (\tv_1\cdot\tv_2)\\
     S_{12}^T &\equiv& 3(\sv_1\cdot\hat R) (\sv_2\cdot\hat R) -
               (\sv_1\cdot\sv_2) \nonumber
\eq
It is well known that the Skyrmion-Skyrmion force contains only the
spin-isospin independent term $V_c$,
$(\sv_1\cdot\sv_2)(\tv_1\cdot\tv_2)$ term $V_s$, and the isovector
tensor term $V_t$.  In the product approximation, we find that the
remaining term in the expansion (\ref{eq:WZexp}) can be neglected.
Once the eq.(\ref{eq:WZexp}) is truncated to the first three terms,
then one can compute $V_c$, $V_s$ and $V_t$ by choosing only three
independent relative orientations of the Skyrmions.

Three quasistatic configurations obtained in the previous section can
be used to deduce the $N-N$ potentials.  The results are summarized in
\Figures{4, 5 and 6}.
\Figure{4} compares the central spin-isospin independent potentials
for the PA and QS configurations.  One sees again that the strong
repulsion in PA has almost gone in QS, although we find no attraction
required for the nuclear binding.  We conclude again that the strong
medium range repulsion seen in PA is an artifact of nonappropriate
configurations.

The spin dependent potentials $V_s$ and $V_t$ are shown in
\Figures{5 and 6}, where one sees that the reduction of the spin-spin
interaction similar to $V_c$, while the tensor two-pion contribution
is stll significant.

\section{Conclusion}

It is promising that the full nuclear force can be understood in the
Skyrme model.  The longest range pion-exchange potential is given
naturally just by placing two Skyrmions.  This is due to the
nonlinearity of the model.  The Skyrmion is  considered as a source of
pions and the classical solutions contain the static part of the pion
cloud in the nucleon.  Thus the static OPE is given in the classical
level as the Coulomb potential between static charges is classical.
Nonadiabatic effects are considered as $1/N_c$ corrections to the
static solution.

In the present study, we show that the medium-range repulsion observed
in the product
approximation
is not real and that the two-pion range Skyrmion-Skyrmion interaction
is indeed small.
The attraction in the medium range remains an open problem yet.
Amado, Walet and Hosaka\cite{Walet}\ recently argued that a careful
treatment of finite $\Nc$ effects may solve the problem.
The showed that mixings of the
$N\Delta$, and
$\Delta\Delta$ components yield
significant attraction.  Their result is comparable to the Reid soft
core potential at $R>2$ fm.
The present calculation does not address such quantal or
higher $1\over\Nc$ effects.
We concentrate on the classical adiabatic potential between Skyrmions.

We would like to stress that the present formalism gives a basis of
a full quantal calculation.
By employing the Dirac quantization method with constraints,
one can study nonadiabatic effects in the two baryon systems.  Indeed,
in ref\cite{paperI}, it was demonstrated in a (1+1) soliton model that
the full recoil meson-exchange interactions is obtained by such an
approach.

\begin{thefigures}{99}

\figitem{1}  Pion isovector field configurations with three different
relative orientations: (a) $A^{\dagger} B =1$, (H), (b) $A^{\dagger} B
= i\tau_x$, (X),
and (c) $A^{\dagger} B = i\tau_z$ (Z).

\figitem{2} The adiabatic Skyrmion-Skyrmion potentials for three
orientations, H, X, and Z.  The dashed lines are the potentials
obtained in the corresponding product approximation.

\figitem{3} The adiabatic Skyrmion-Skyrmion potentials after
subtracting the one-pion exchange term.  See captions for Fig.~2.

\figitem{4} The central part of the nucleon-nucleon potential.  The
dashed curve is that in the product approximation.

\figitem{5} The spin-spin and the tensor parts of the nucleon-nucleon
potential.

\figitem{6} The spin-spin and the tensor parts of the nucleon-nucleon
potential after the one-pion exchange part subtracted.

\end{thefigures}


\begin{thebibliography}{99}

\bibitem{Sky}  T.H.R.~Skyrme, Proc.\ Roy.\ Soc.\ London, \pnum A260,127,61.;
               \NP 31,5,61..
\bibitem{ANW}  G.S.\ Adkins, C.R.\ Nappi and E.\ Witten,
                               \NP B228,552,83.;
         G.S.\ Adkins and C.R.\ Nappi, \NP B233,109,84.;
         A.D.\ Jackson and M.\ Rho, \PRL 51,751,83..
\bibitem{Zahed}         I.\ Zahed and G.E.\ Brown,  Physics Reports
                \pnum 142,1,86..
\bibitem{Jackson}  A.\ Jackson, A.D.\ Jackson, and V.\ Pasquier,
                      \NP A432,567,85.;
\bibitem{VM}     R.\ Vinh \ Mau, M.\ Lacombe, B.\ Loiseau,
              W. N.\ Cottingham, and P.\ Lisboa, \PL 150,259,85..
\bibitem{Oka}  \MO, K.F.\ Liu\ and H.\ Yu, \PRD 34,1575,86.;
         \MO, \NP A463,247c,87.; \PRC 36,720,87..
\bibitem{OH}   M.~Oka and A.~Hosaka, Annual Rev.\ of Nucl.\ Part.\
Sci.,    \pnum 42,333,92..
\bibitem{Wal}  T.~S.~Walhout and J.~Wambach, \PRL 67,314,91..
\bibitem{AM}   M.~F.~Atiyah and N.~S.~Manton, \PL 222,438,89.;
                   N.~S.~Manton, ``Proceedings of the LMS Symposium:
                   Geometry of Low Dimensional Manifolds'' (Durham,
                    1989).
\bibitem{AM-hosaka}     A.\ Hosaka, S.M.\ Griffies, \MO\ and \AM, \PL
                            251,1,90.;
                   A.~Hosaka, M.~Oka and R.~D.~Amado, \NP
                      A530,507,91..
\bibitem{PRL}  M.\ Oka, \PRL 66,1019,91..
\bibitem{donuts}  J.J.M.\ Verbaarschot, \PL 195,235,87.;
         E.\ Braaten and L.\ Carson, \PRD 38,3525,88..
\bibitem{Manton}  N.S.\ Manton, \PRL 60,1916,88..
\bibitem{paperI}  R.D.\ Amado, H.\ Liu, M.\ Oka and E.\ Wong, \PRL
63,852,89.;
         M.\ Oka, H.\ Liu and R.D.\ Amado,   \PRC 39,2317,89..
\bibitem{Dirac}   P.A.M.~Dirac, ``Lectures in Quantum Mechanics'',
                   Belfer Grad.\ Sch.\ of Sci., Yeshiva Univ.\ (1964).
\bibitem{OBBA}    R.~D.~Amado, R.~Bijker and M.~Oka, \PRL 58,654,87.;
                  \MO, R.\ Bijker, A.\ Bulgac and R.D.~Amado, \PRC
                    36,1727,87..
\bibitem{Walet}  N.R.~Walet, R.D.~Amado and A.~Hosaka, \PRL
                 68,3849,92.;
                 N.R.~Walet and  R.D.~Amado, University of
                       Pennsylvania preprint, UPR-0124MT.
\end{thebibliography}
\end{document}